\documentclass[12pt]{article}
\usepackage{subfloat}
\pdfoutput = 1
\usepackage{graphics}
\usepackage{graphicx} 
\usepackage{amsmath}
\usepackage{amsfonts}   
\usepackage{amssymb}
\usepackage{multirow}
\usepackage{hhline}
\usepackage{color}
\usepackage{comment}
\begin{document}
\date{Today}
\title{{\bf{\Large Effects of massive gravity on $s$-wave holographic superconductor}}}
\author{ {\bf {\normalsize Diganta Parai}$^{a}$
\thanks{digantaparai007@gmail.com}},\,
{\bf {\normalsize Suchetana Pal}$^{a}
$\thanks{suchetanapal92@gmail.com, sp15rs004@iiserkol.ac.in}},\,
{\bf {\normalsize Sunandan Gangopadhyay}$^{b}$
\thanks{sunandan.gangopadhyay@gmail.com, sunandan.gangopadhyay@bose.res.in}}\\
$^{a}$ {\normalsize Indian Institute of Science Education and Research Kolkata}\\{\normalsize Mohanpur, Nadia 741246, India}\\[0.2cm] 
$^{b}$ {\normalsize Department of Theoretical Sciences } \\{\normalsize S.N. Bose National Centre for Basic Sciences }\\{\normalsize JD Block, 
Sector III, Salt Lake, Kolkata 700106, India}\\[0.2cm]
}
\date{}

\maketitle
\begin{abstract}
 Analytical investigation of the properties of $s$-wave holographic superconductors in the background of a massive gravity theory in the probe limit has been carried out employing the Sturm-Liouville  eigenvalue method . We obtain the analytical expression for the relation between the critical temperature and the charge density. We also obtain the expression for the condensation operator and value of the critical exponent. Our findings show that as we increase the massive gravity couplings the critical temperature increases and the condensate decreases. More precisely we observe that the presence of massive graviton increases the critical temperature compared to the superconductors in Einstein gravity at some point if we keep on increasing the coupling constants. We also obtain the frequency dependence of conductivity by solving analytically the wave equation for electromagnetic perturbations. From the real part of the conductivity, we finally estimate the energy band gap.     
\end{abstract}
\vskip 1cm
\section{Introduction}

\noindent The AdS/CFT correspondence also known as gauge/gravity duality, originated from string theory  provides the correspondence between weakly coupled gravity theory in a $(d+1)$ dimensional $AdS$ spacetime and a strongly coupled conformal field theory (CFT) residing on its $d$-dimensional boundary \cite{adscft1}-\cite{adscft4}. In recent years, the AdS/CFT correspondence has been used extensively to investigate several properties of strongly coupled systems. The reason behind this is as follows. The weakly coupled gravity system can be tackled by perturbation theory. Then using this correspondence one can get a picture of some of the properties of the corresponding strongly coupled field theoretic system living on the boundary of $AdS$ spacetime.

\noindent The studies conducted to understand high $T_{c}$ superconductors in the framework of usual Maxwell \cite{adscft5}-\cite{horowitz} as well as Born infeld electrodynamics  \cite{hs19}-\cite{dpdgsg} has been carried out extensively. The  holographic  description of a $s$-wave superconductor can be described by a charged black hole, minimally coupled to a complex scalar field which manifests local spontaneous $U(1)$ symmetry breaking. The formation of scalar hair below certain critical temperature ($T_{c}$) indicates the onset of a condensation in the dual CFT. Another key aspect of holographic superconductors is their response to an external magnetic field known as the Meissner effect has been carried out in the framework of usual Maxwell electromagnetic theory as well as in power Maxwell electrodynamics \cite{su, sup}. Examples of black holes exhibiting $p$-wave superconductivity have been observed \cite{ss}-\cite{spdg}. Holographic insulator to superconducor phase transition has also been studied in \cite{dg1}-\cite{dg3}.

\noindent As mentioned earlier, the holographic superconductor has been modeled successfully by Einstein gravity coupled to a matter field. But it is always interesting to explore such systems considering the graviton to be massive. Massive gravity is a modification of Einstein gravity in the infrared(IR) region which includes the mass terms for graviton. The theoretical construction of massive gravity was first attempted by Fierz and Pauli \cite{mw}. Unfortunately, this attempt suffered
discrepancies both in linear and nonlinear levels known as the van Dam-Veltman-Zakharov (vDVZ) discontinuity \cite{hm,vp} and the Boulware-Deser (BD) ghost \cite{ds} problem respectively. In 2010, de Rham, Gabadadze and Tolley proposed a successful nonlinear massive gravity theory which eliminated both of these problems \cite{cg,cga}.

\noindent Considering massive gravity $s$-wave holographic superconductor was studied numerically \cite{hj,ry} and $p$-wave holographic superconductor was also investigated analytically \cite{CHNam}.
In this paper we have constructed a $s$-wave holographic superconductor model taking into account the massive gravity and analytically studied the role and physical impact of the graviton mass terms on the properties of $s$-wave holographic superconductor.

\noindent This paper is organized as follows. In Section 2, we have introduced a $s$-wave holographic superconductor model in the probe limit in the black brane geometry background considering massive graviton. In Section 3, the Sturm-Liouville method has been used to obtain the critical temperature-charge density relationship analytically and the behaviours of the critical temperature have been investigated in presence of massive graviton. In Section 4, we have determined an analytical expression of the condensation operator and studied its behaviour in the presence of the mass of graviton. In section 5 we obtain an expression of optical conductivity for our model with a particular boundary condition and compute the band gap energy. Finally, in Section 6 we have concluded.

\section{Basic set up}
\noindent In order to construct a s-wave holographic superconductor model considering massive gravity, we will start with the following action \cite{CHNam}
\begin{eqnarray}
S=\int d^{d}x \sqrt{-g}\Big[R-2\Lambda+m_{g}^2\sum_{i=1}^{4}c_{i}\mathcal{U}_{i}(g,f) +\mathcal{L}_{matter}\Big] ,
\label{1}
\end{eqnarray} 
\noindent where $R$ is the scalar curvature of the space time, $\Lambda=-(d-1)(d-2)/(2L^2)$ is the cosmological constant, $L$ is the $AdS$ radius, $m_{g}$ is the mass of graviton, $c_{i}$ are the coupling parameters, $f$ is the reference metric which is a fixed symmetric tensor, $\mathcal{U}_{i}$ can be written as symmetric polynomials in terms of the eigenvalues of the $d\times d$ matrix $\mathcal{K}^{\mu }_{~\nu }=\sqrt{g^{\mu \lambda }f_{\lambda \nu }}$  
\begin{eqnarray}
\mathcal{U}_1=[\mathcal{K}],~~\mathcal{U}_2=[\mathcal{K}]^2-[\mathcal{K}^2],~~\mathcal{U}_3=[\mathcal{K}]^3-3[\mathcal{K}][\mathcal{K}^2]+2[\mathcal{K}^3],\nonumber\\
\mathcal{U}_4=[\mathcal{K}]^4-6[\mathcal{K}]^2[\mathcal{K}^2]+8[\mathcal{K}][\mathcal{K}^3]+3[\mathcal{K}^2]^2-6[\mathcal{K}^4] ,
\label{2}
\end{eqnarray}

\noindent where rectangular
brackets denote traces: $[\mathcal{K}]=\mathcal{K}^{\mu}_{~\mu}$.

\noindent The matter Lagrangian density denoted by $\mathcal{L}_{matter}$ reads
\begin{eqnarray}
\mathcal{L}_{matter}= - \frac{1}{4}F^{\mu \nu} F_{\mu \nu} - (D_{\mu}\psi)^{*} D^{\mu}\psi-m^2 \psi^{*}\psi , 
\label{3}
\end{eqnarray}
where $F_{\mu \nu}=\partial_{\mu}A_{\nu}-\partial_{\nu}A_{\mu}$ is the field strength tensor, 
$D_{\mu}=(\partial_{\mu}-iqA_{\mu})$ is the covariant derivative with $m$ and $q$ being
the mass and charge of the complex scalar
field $\psi$.\\
\noindent Now let us consider the following $d$ dimentional plane symmetric black hole solution in $AdS$ spacetime with the metric ansatz
\begin{eqnarray}
ds^2=-f(r)dt^2+\frac{1}{f(r)}dr^2+ r^2 h_{ij} dx^{i} dx^{j},
\label{4}
\end{eqnarray}
where $h_{ij} dx^{i} dx^{j}$ denotes the line element of a $(d-2)$ dimensional hypersurface with zero curvature. Following \cite{v}, we consider the reference metric in the form as $f_{\mu\nu}=diag(0,0,h_{ij})$. The ansatz for the gauge field and the scalar field is now chosen to be
\begin{eqnarray}
A_{\mu} = (\phi(r),0,0,0)~,~\psi=\psi(r).
\label{5}
\end{eqnarray}
\noindent Varying the action (\ref{1}) with respect to the chosen ansatz in the probe limit \footnote{ Working in probe limit signifies we neglect the efect of back-reaction of the matter field on the metric. In this limit we take the charge $q$ to be large and rescale $\psi\rightarrow \frac{\psi}{q}$ and $\phi\rightarrow \frac{\phi}{q}$.}, we obtain the following equations of motion 

\begin{eqnarray}
\frac{d-2}{2r^4}\big[r^2(d-3)f+r^3f^{'}\big]-\frac{(d-1)(d-2)}{2L^2}-\frac{m_{g}^2}{2}\Big[\frac{(d-2)c_1}{r}+\frac{(d-2)(d-3)c_2}{r^2}\nonumber\\
+\frac{(d-2)(d-3)(d-4)c_3}{r^3}+\frac{(d-2)(d-3)(d-4)(d-5)c_4}{r^4} \Big]=0
\label{6}
\end{eqnarray}
\begin{eqnarray}
\phi^{\prime \prime}(r) + \frac{d-2}{r}\phi^{\prime}(r) - \frac{2 q^2  \psi^{2}(r)}{f(r)} \phi(r)= 0
\label{7}
\end{eqnarray}
\begin{eqnarray}
\psi^{\prime \prime}(r) + \left(\frac{d-2}{r}+ \frac{f^{\prime}(r)}{f(r)}\right)\psi^{\prime}(r) + \left(\frac{q^2 \phi^{2}(r)}{f(r)^2}- \frac{m^{2}}{f(r)}\right)\psi(r) = 0 ,
\label{8}
\end{eqnarray}
where prime denotes derivative with respect to $r$.
 Without any loss of generality one can choose $q=1$  since the rescalings $\psi\rightarrow \psi/q$ and $\phi\rightarrow \phi/q$ can be performed. By solving eq.(\ref{6}) we get the following metric function $f(r)$ \cite{CHNam}
 \begin{eqnarray}
 f(r)=\frac{r^2}{L^2}\Bigg(1-\frac{r_{+}^{d-1}}{r^{d-1}}\Bigg)+m_{g}^2\Bigg[\frac{c_1r}{d-2}\Bigg(1-\frac{r_{+}^{d-2}}{r^{d-2}}\Bigg)+c_2\Bigg(1-\frac{r_{+}^{d-3}}{r^{d-3}}\Bigg)\nonumber\\
 +\frac{(d-3)c_3}{r}\Bigg(1-\frac{r_{+}^{d-4}}{r^{d-4}}\Bigg)+\frac{(d-3)(d-4)c_4}{r^2}\Bigg(1-\frac{r_{+}^{d-5}}{r^{d-5}}\Bigg)\Bigg] ,
 \label{9}
 \end{eqnarray}
where $r_{+}$ is the event horizon radius. The temperature associated to the  boundary field theory can be identified as the Hawking temperature of the black hole and is given by 
\begin{eqnarray}
T=\frac{(d-1)r_{+}}{4\pi L^2}+\frac{m_{g}^2}{4\pi}\Bigg[c_1+\frac{(d-3)c_2}{r_{+}}+\frac{(d-3)(d-4)c_3}{r_{+}^2}+\frac{(d-3)(d-4)(d-5)c_4}{r_{+}^3}\Bigg].
\label{10}
\end{eqnarray}
\noindent To solve the above non-linear differential equations (\ref{7}) and (\ref{8}), we must set physically acceptable boundary condition for  $\phi(r)$ and $\psi(r)$ . For regularization of the matter filed, one requires  $\phi(r_+)=0$ and $\psi(r_{+})$ to be finite at the horizon. The asymptotic behaviour ($r\rightarrow\infty$) of the fields can be written as 

\begin{eqnarray}
\phi(r)=\mu-\frac{\rho}{r^{d-3}}
\label{11}
\end{eqnarray}
\begin{eqnarray}
\psi(r)=\frac{\psi_{-}}{r^{\Delta_{-}}}+\frac{\psi_{+}}{r^{\Delta_{+}}},
\label{12}
\end{eqnarray}
where
\begin{eqnarray}
\Delta_{\pm}&=&\frac{(d-1)\pm\sqrt{(d-1)^2+4m^2 }}{2} .
\label{13}
\end{eqnarray}
\noindent $\mu$ and $\rho$ in eq.(\ref{11}) can be interpreted  as the chemical potential and charge density respectively, and the coefficients $\psi_{-}$ and $\psi_{+}$ correspond to the vacuum expectation values of the condensation operator $\mathcal{O}_{\Delta}$ dual to the boundary field theory. As we can impose the condition that either $\psi_{-}$ or $\psi_{+}$ vanishes at the asympotic $AdS$ region to make the theory stable, we also set $\psi_{-}=0$ and $\psi_{+}=\langle \mathcal{O}_{\Delta_+}\rangle$.
$\Delta_{\pm}$ is the conformal dimension of the condensation operator.

\noindent Under the change of coordinates $z=\frac{r_{+}}{r}$, the field equations (\ref{7},\ref{8}) become
\begin{eqnarray}
\phi^{\prime \prime}(z) - \frac{d-4}{z} \phi^{\prime}(z)  - \frac{2r^2_{+} \psi^{2}(z)}{f(z) z^4}\phi(z) = 0 
\label{14}
\end{eqnarray}

\begin{eqnarray}
\psi^{\prime \prime}(z) + \left(\frac{f^{\prime}(z)}{f(z)} - \frac{d-4}{z}\right)\psi^{\prime}(z) + \frac{r^2_{+}}{z^4} \left(\frac{\phi^{2}(z)}{f(z)^2}- \frac{m^{2}}{f(z)}\right)\psi(z)=0 ,
\label{15}
\end{eqnarray}

\noindent where prime now denotes derivative with respect to $z$. These equations are to be solved in the interval $(0, 1)$, where $z=1$ is the horizon and $z=0$ is the boundary.
The boundary condition $\phi(r_+)=0$ now changes to $\phi(z=1)=0$.
$f(z)$ is given by

\begin{eqnarray}
f(z)=\frac{r_{+}^2}{L^2z^2}\Big(1-z^{d-1}\Big)+m_{g}^2\Bigg[\frac{c_1r_{+}}{(d-2)z}\Big(1-z^{d-2}\Big)+c_{2}\Big(1-z^{d-3}\Big)\nonumber\\
+\frac{(d-3)c_3z}{r_{+}}\Big(1-z^{d-4}\Big)+\frac{(d-3)(d-4)c_{4}z^2}{r_{+}^2}\Big(1-z^{d-5}\Big)\Bigg].
\label{16}
\end{eqnarray}
\noindent In the following sections, we aim to solve these equations in order to investigate the effects of massive graviton on our $s$-wave holographic superconductor model. For convenience we have fixed $L=1$.  
\section{Relation between critical temperature($T_{c}$) and charge density($\rho$)}

\noindent We know at $T\geq T_{c}$, the matter field $\psi(z)$ must vanish. At $T=T_{c}$, we substitute $ \psi(z)=0$ in eq.(\ref{14}), and obtain

\begin{eqnarray}
\phi^{\prime \prime}(z) -\frac{d-4}{z}\phi^{\prime}(z)=0.
\label{17}
\end{eqnarray}

\noindent Using the boundary condition of $\phi(z)$ from eq.(\ref{11}), we solve the above equation and get

\begin{equation}
\phi(z)=\lambda r_{+(c)}\big(1-z^{d-3}\big)\equiv \lambda r_{+(c)}{\xi}(z),
\label{18}
\end{equation}
where $r_{+(c)}$ is the horizon radius at $T=T_{c}$ and 
\begin{equation}
    \lambda =\frac{\rho }{r_{+(c)}^{d-2}}.
    \label{19}
\end{equation}

\noindent Let us now consider the variable change $\bar{c_{n}} \equiv \frac{c_{n}}{r_{+(c)}^{n}}$ for $n=1,2,3,4$. Now at $T=T_{c}$ ,the metric (\ref{16}) reads
\begin{eqnarray}
f(z)=\frac{r_{+(c)}^2}{z^2}\Bigg[1-z^{d-1}+m_{g}^2\Bigg\{ \frac{\bar{c_{1}}z(1-z^{d-2})}{(d-2)}+\bar{c_2}z^2(1-z^{d-3})\nonumber\\
+(d-3)\bar{c_3}z^3(1-z^{d-4})+(d-3)(d-4)\bar{c_4}z^4(1-z^{d-5})\Bigg\}\Bigg] \equiv \frac{r_{+(c)}^2}{z^2} g(z).
\label{22}
\end{eqnarray}
\noindent We define 
\begin{eqnarray}
g_{0}(z)&=&1-z^{d-1}\nonumber\\
g_{1}(z)&=&m_{g}^2\Bigg\{ \frac{\bar{c_{1}}z(1-z^{d-2})}{(d-2)}+\bar{c_2}z^2(1-z^{d-3})
+(d-3)\bar{c_3}z^3(1-z^{d-4})\nonumber\\
&+&(d-3)(d-4)\bar{c_4}z^4(1-z^{d-5})\Bigg\}
\label{22a}
\end{eqnarray}
\noindent Next we substitute $\phi(z)$ from eq.(\ref{18}) and $f(z)$ from eq.(\ref{22}) in eq.(\ref{15}).With this we find that and near the critical temperature $T\rightarrow T_{c}$, the scalar field equation for $\psi(z)$ takes the form

\begin{eqnarray}
\psi^{\prime\prime}(z)+\left(\frac{g^{\prime}(z)}{g(z)}-\frac{d-2}{z}\right )\psi^{\prime}(z)+\left(\frac{\lambda^2\xi^2(z)}{g^2(z)}-\frac{m^2}{z^{2}g(z)} \right )\psi(z)=0.
\label{23}
\end{eqnarray}
\noindent Keeping the behaviour at the boundary ($z\rightarrow 0$) in mind, we define \cite{siop} 

\begin{equation}
    \psi(z)=\frac{\langle \mathcal{O}_{\Delta_+}\rangle}{r_{+}^{\Delta _{+}}}z^{\Delta _{+}}F(z).
\label{24}    
\end{equation}

\noindent To make this  expression consistent with eq.(\ref{12}), $F$ is normalized to $F(0)=1$. Now from eq.(\ref{23}), we obtain

\begin{eqnarray}
F^{\prime\prime}(z)&+&\Bigg\{\frac{2\Delta _{+}-d+2}{z}+\frac{g^{\prime}(z)}{g(z)}\Bigg\}F^{\prime}(z)\nonumber\\&+&\Bigg\{\frac{\Delta _{+}}{z}\left(\frac{\Delta _{+}-d+1}{z}+\frac{g^{\prime}(z)}{g(z)}\right )-\frac{m^2}{z^2g(z)}\Bigg\}F(z)+\lambda ^{2}\frac{\xi ^2(z)}{g^2(z)}F(z)=0
\label{25}
\end{eqnarray}

\noindent to be solved subject to the boundary condition $F^{\prime}(0)=0$.\\

\noindent Our next objective is to find out the value of $\lambda$. To achieve that we rewrite eq.(\ref{25}) in the Sturm-Liouville  form
\begin{equation}
    \frac{\mathrm{d} }{\mathrm{d} z}\Big\{p(z)F^{\prime}(z)\Big\}+q(z)F(z)+\lambda ^2r(z)F(z)=0
    \label{26}
\end{equation}

with
\begin{eqnarray}
&p(z)&=z^{2\Delta _{+}-d+2}g(z)\nonumber\\
&q(z)&=p(z)\Bigg\{\frac{\Delta _{+}}{z}\left(\frac{\Delta _{+}-d+1}{z}+\frac{g^{\prime}(z)}{g(z)} \right )-\frac{m^2}{z^2g(z)}\Bigg\}\nonumber\\
&r(z)&=\frac{p(z)\xi^2(z)}{g^2(z)}~.    
\label{27}
\end{eqnarray}

\noindent Now the eigen value
$\lambda^2$ of this Sturm Liouville problem can be written in following form,  extremization of which leads to eq.(\ref{26})\cite{siop}

\begin{eqnarray}
\lambda^{2}=\frac{\int_{0}^{1}dz\big\{p(z)[F^{\prime}(z)]^{2}-q(z)[F(z)]^{2}\big\}}{\int_{0}^{1}dz r(z)[F(z)]^{2}}~.
\label{28}
\end{eqnarray}
To estimate $\lambda^2$, let us consider the following trial solution which satisfies the previously mentioned boundary conditions $F(0)=1$ and $F^{\prime}(0)=0$ 
\begin{eqnarray}
F=F_{\alpha}(z)=1-\alpha z^2.
\label{29}
\end{eqnarray}

\noindent Minimizing eq.(\ref{28}) with respect to $\alpha$ we obtain certain values of $\lambda$ denoted by $\lambda_{\alpha}$ depending on different coupling coefficients $\bar{c_{n}}$. These values of $\lambda_{\alpha}$ are listed in Table [\ref{tab1}] and Table [\ref{tab2}].

\noindent Now we proceed to obtain the relation between the critical temperature and charge density. To do this we start from eq.(\ref{10}) and substitute $r_{+(c)}$ from eq.(\ref{19}). This yields

\begin{eqnarray}
T_{c}=\frac{r_{+(c)}}{4\pi}\left [ (d-1)+m_{g}^2\left \{ \bar{c_{1}}+(d-3)\bar{c_{2}}+(d-3)(d-4)\bar{c_{3}}+(d-3)(d-4)(d-5)\bar{c_{4}} \right \} \right ]\nonumber\\
=\frac{1}{4\pi}\left [ (d-1)+m_{g}^2\left \{ \bar{c_{1}}+(d-3)\bar{c_{2}}+(d-3)(d-4)\bar{c_{3}}+(d-3)(d-4)(d-5)\bar{c_{4}} \right \} \right ]\left(\frac{\rho}{\lambda_{\alpha}}\right)^\frac{1}{d-2}~.
\label{30}
\end{eqnarray}

\noindent It is clear from eq.(\ref{30}) that the critical temperature $T_{c}$ depends on the coupling coefficients of massive gravity.  Note that in four dimensions only the couplings $c_{1}$, $c_{2}$ affect the critical temperature. whereas, the couplings $c_{3}$ and $c_{4}$ modify the critical temperature only when the space time dimension is higher than four. To observe the effect of massive gravity couplings on critical temperature, we have listed the ratio $T_{c}/\rho^{\frac{1}{d-2}}$ for different massive gravity couplings in Table-\ref{tab1} for $d=4$  and in Table-\ref{tab2} for $d=5$.

\noindent Now let us consider the particular cases when all massive gravity couplings coefficients are zero (Einstein gravity). For $d=5$ and $m^2=-3$ critical temperature turns out to be $T_{c}=0.1962 \rho^{\frac{1}{3}}$ which is in very good agreement with the numerical result $T_{c}=0.1980\rho^{\frac{1}{3}}$ \cite{gm}. For $m^2=0$, $T_{c}=0.0844 \rho^{\frac{1}{2}}$ for $d=4$ and $T_{c}=0.1676 \rho^{\frac{1}{3}}$ at $d=5$.

\noindent We observe that the increase in the massive gravity couplings increases the critical temperature. From Table-\ref{tab1} and Table-\ref{tab2}, we see that there exists a critical value for each massive gravity coupling above which the critical temperature in massive gravity is higher than that in Einstein gravity and hence the high temperature superconductors can be achieved in the framework of massive gravity. 

\begin{table}[h!]
\caption{The analytical results of the ratio $\left(\frac{T_{c}}{\rho^{\frac{1}{2}}}\right)$ for different values of $\bar{c_{1}}$ and $\bar{c_{2}}$ at $d=4$, $m_{g}=1$, $m^2=0$.}
    \centering
    \begin{tabular}{|c|c|c|c|c|c|}
    \hline
    $\bar{c_{1}}=1$&$\lambda_{\alpha}$&$\frac{T_{c}}{\rho^{\frac{1}{2}}}$ &$\bar{c_{2}}=2$&$\lambda_{\alpha}$ &$\frac{T_{c}}{\rho^{\frac{1}{2}}}$ \\
    \hhline{-~~-~~}
    $\bar{c_{2}}$& & &$\bar{c_{1}}$& & \\
    \hline
    -1 & 8.502 & 0.0819 & -2 & 7.004 & 0.0902 \\
    \hline
    -0.5 & 9.115 & 0.0923 & -1 & 8.702 & 0.1079 \\
    \hline
    0 & 9.718 & 0.1021 & 0 & 10.388 & 0.1234 \\
    \hline
    0.5 & 10.313 & 0.1115 & 1 & 12.065 & 0.1374 \\
    \hline
    1 & 10.9019 & 0.1205 & 2 & 13.738 & 0.1503\\
    \hline
    \end{tabular}
    \label{tab1}
\end{table}

\begin{table}[h!]
\caption{The analytical results of the ratio $\left(\frac{T_{c}}{\rho^{\frac{1}{3}}}\right)$ for different values of $\bar{c_{1}}$, $\bar{c_{2}}$ and $\bar{c_{3}}$ at $d=5$, $m_{g}=1$, $m^2=0$.}
    \centering
    \begin{tabular}{|c|c|c|c|c|c|c|c|c|}
    \hline
    $\bar{c_{2}}=-\bar{c_{3}}=1$& $\lambda_{\alpha}$ & $\frac{T_{c}}{\rho^{\frac{1}{3}}}$&
    $\bar{c_{3}}=-\bar{c_{1}}=1$& $\lambda_{\alpha}$ &
    $\frac{T_{c}}{\rho^{\frac{1}{3}}}$&
    $\bar{c_{1}}=-\bar{c_{2}}=1$& $\lambda_{\alpha}$ &
    $\frac{T_{c}}{\rho^{\frac{1}{3}}}$ \\
    \hhline{-~~-~~-~~}
    $\bar{c_{1}}$ & & & $\bar{c_{2}}$ & & & $\bar{c_{3}}$ & & \\
    \hline
    -1 & 6.038 & 0.1311 & -1 & 5.264 & 0.1372 & -1 & 4.555 & 0.048\\
    \hline
    -0.5 & 6.640 & 0.1482 & -0.5 & 6.150 & 0.1737 & -0.5 & 5.481 & 0.0903\\
    \hline
    0 & 7.238 & 0.1646 & 0 & 7.015 & 0.2078 & 0 & 6.249 & 0.1296\\
    \hline
    0.5 & 7.832 & 0.1803 & 0.5 & 7.866 & 0.2401 & 0.5 & 6.960 & 0.1667\\
    \hline
    1 & 8.424 & 0.1955 & 1 & 8.709 & 0.2707 & 1 & 7.641 & 0.2020\\
    \hline
    \end{tabular}
    
    \label{tab2}
\end{table}

\section{Condensation operator and critical exponent }
In this section, we aim to investigate relation between condensation operator $\langle J\rangle$ and critical temperature $T_{c}$. Let us start with equation(\ref{14}) for $\phi (z)$  near the critical temperature $T_{c}$ 
\begin{eqnarray}
\phi^{\prime \prime}(z) - \frac{d-4}{z} \phi^{\prime}(z) = \frac{\langle \mathcal{O}_{\Delta_+}\rangle ^{2}}{r^{2}_{+}} \mathcal{B}(z)\phi (z),
\label{31} 
\end{eqnarray}
where $\mathcal{B}(z)= \frac{2 z^{2\Delta_{+} -4}}{r_{+}^{2\Delta_{+} -4}}\frac{F^{2}(z)}{f(z)}$.
\noindent  We may now expand $\phi(z)$ in terms of the small parameter $\frac{\langle \mathcal{O}_{\Delta_+} \rangle ^2}{r^{2}_{+}}$ as 
\begin{eqnarray}
\frac{\phi(z)}{r_{+}} = \lambda \xi(z)+ \frac{\langle \mathcal{O}_{\Delta_+}\rangle ^2}{r^{2}_{+}} \zeta (z),
\label{32} 
\end{eqnarray}
where $\zeta(z)$ is the correction term and $\zeta (1)= 0 =\zeta^{\prime}(1).$\\
Substituting $\phi(z)$ from eq.(\ref{32}) in eq.(\ref{31})and comparing the coefficient of $\frac{\langle \mathcal{O}_{\Delta_+}\rangle ^2}{r^{2}_{+}}$ on both sides of this equation, we get the equation for the correction $\zeta(z)$ near the critical temperature
\begin{eqnarray}
\zeta^{\prime \prime}(z) - \frac{d-4}{z}\zeta^{\prime}(z) = \lambda \frac{2 z^{2\Delta_{+} -4}}{r_{+}^{2\Delta_{+} -4}}\frac{F^{2}(z)}{f(z)}\xi (z)~.
\label{33}
\end{eqnarray}

\noindent By multiplying $z^{-(d-4)} $ on the both sides of eq.(\ref{33}), we get
\begin{eqnarray}
\frac{d}{dz}\Big\{z^{-(d-4)}  \zeta^{\prime}(z) \Big\} = \lambda \frac{2 z^{2\Delta_{+} -d+2}}{r_{+}^{2\Delta_{+} -2}}\frac{ F^{2}(z)}{g(z)}\xi(z). 
\label{34}
\end{eqnarray}
Using the boundary conditions of $\zeta(z)$, we integrate the above equation between the limits $z=0$ and $z=1$. This leads to
\begin{eqnarray}
\frac{\zeta^{\prime}(z)}{z^{d-4}}\mid_{z\rightarrow 0} = -\frac{\lambda}{r^{2\Delta_{+} -2}_{+}} \mathcal{A}_{1},
\label{35}
\end{eqnarray}
where $\mathcal{A}_{1} = \int^{1}_{0} dz \frac{2 z^{2\Delta_{+}-d+2} F^{2}(z)}{g(z)} \xi(z) $.\\

\noindent We may get the asymptotic behaviour ($z\rightarrow0$) of $\phi(z)$ from eq.(\ref{32}). By comparing it with 
eq.(\ref{11}), we obtain 
\begin{eqnarray}
\mu -\frac{\rho}{r^{d-3}_{+}}z^{d-3} = \lambda r_{+} (1-z^{d-3})+ \frac{\langle \mathcal{O}_{\Delta_+}\rangle ^2}{r_{+}} \left\{\zeta(0)+z\zeta^{\prime}(0)+......+\frac{\zeta^{d-3}(0)}{(d-3)!} z^{d-3}+....\right\}.
\label{36}
\end{eqnarray}
Comparing the coefficient of $z^{d-3}$ on both sides of this equation, we get
\begin{eqnarray}
-\frac{\rho}{r^{d-3}_{+}} = -\lambda r_{+} + \frac{\langle \mathcal{O}_{\Delta_+}\rangle ^2}{r_{+}}\frac{\zeta^{d-3}(0)}{(d-3)!}
\label{37}
\end{eqnarray}
 together with $\zeta^{\prime}(0)=\zeta^{\prime\prime}(0)=.......\zeta^{d-4}(0)=0.$

\noindent Now we note that $\zeta^{\prime}(z)$ and $(d-3)$-th derivative of $\zeta(z)$ are related by
\begin{eqnarray}
\frac{\zeta^{(d-3)}(z=0)}{(d-4)!} = \frac{\zeta^{\prime}(z)}{z^{d-4}}|_{z\rightarrow 0}.
\label{38}
\end{eqnarray}
From eq.(s)(\ref{35}, \ref{38}) and eq.(\ref{37}), we obtain the relation between the charge density $\rho$ and the condensation operator $\langle \mathcal{O}_{\Delta_+}\rangle$ 
\begin{eqnarray}
\frac{\rho}{r^{d-2}_{+}} = \lambda\left[1 + \frac{\langle \mathcal{O}_{\Delta_+}\rangle ^2}{r^{2\Delta_{+}}_{+}}\frac{\mathcal{A}_{1}}{(d-3)} \right].
\label{39}
\end{eqnarray}
\noindent Substituting $\lambda$ from eq.(\ref{19}) and using the definition of $T$ from eq.(\ref{39}) we obtain 
\begin{eqnarray}
\langle \mathcal{O}_{\Delta_+}\rangle ^2=\frac{(d-3)(4\pi T)^{2\Delta _{+}}}{\Big(d-1+m_{g}^2\left \{ \bar{c_{1}}+(d-3)\bar{c_{2}}+(d-3)(d-4)\bar{c_{3}}+(d-3)(d-4)(d-5)\bar{c_{4}} \right \}\Big)^{2\Delta _{+}}\mathcal{A}_{1}}\left(\frac{T_{c}}{T}\right)^{d-2}
\nonumber\\
\times\left[1- \left(\frac{T}{T_{c}}\right)^{d-2} \right].
\label{40}
\end{eqnarray}
Now we finally obtain the relation between the condensation operator and the critical temperature in $d$-dimension for $T\rightarrow T_c$
\begin{eqnarray}
\langle \mathcal{O}_{\Delta_+}\rangle = \beta T^{\Delta_{+}}_{c} \sqrt{1-\frac{T}{T_{c}}}
\label{41}
\end{eqnarray}
where $\beta = \sqrt{\frac{(d-3)(d-2)}{\mathcal{A}_{1}}}\left [ \frac{4\pi}{d-1+m_{g}^2\left \{ \bar{c_{1}}+(d-3)\bar{c_{2}}+(d-3)(d-4)\bar{c_{3}}+(d-3)(d-4)(d-5)\bar{c_{4}} \right \}} \right ]^{\Delta _{+}}~.$

\noindent   At $d=5$ and $m^2=-3$, for all massive gravity couplings  set to zero, the the value of $\beta$ is 238.905  which is in very good agreement with the exact result $\beta= 238.958$ \cite{hs24}.

\noindent To observe the effect of massive gravity couplings on the condensate value  we take the value of $\beta$ at different massive gravity couplings in Table-\ref{tab3} for $d=4$ and $m^{2}=0$. We see that condensate value gets smaller on increasing the massive gravity couplings. When all massive gravity couplings are zero (Einstein gravity) and $m^2=0$, we get $\beta=537.305$ for $d=4$. From Table [\ref{tab3}], it can be seen that if we  increase massive gravity couplings it makes $\beta$ smaller than that in Einstein gravity.  This suggests that after a critical value, for each massive gravity coupling  the condensate gets easier to form in massive gravity than Einstein gravity, which is consistent with the behaviour of the critical temperature as indicated in the previous section. 

\begin{table}[h!]
\caption{Analytical results for $\beta$ with different values of $\bar{c_{1}}$ and $\bar{c_{2}}$ at $d=4$, $m_{g}=1$, $m^2=0$. }
    \centering
     \begin{tabular}{|c|c|c|c|}
    \hline
    $\bar{c_{1}}=1$&$\beta$ &$\bar{c_{2}}=2$ &$\beta$ \\
    \hhline{-~-~}
    $\bar{c_{2}}$& &$\bar{c_{1}}$& \\
    \hline
    -1 & 546.34 & -2 & 516.681 \\
    \hline
    -0.5 & 366.565 & -1 & 250.686 \\
    \hline
    0 & 259.193 & 0 & 142.941 \\
    \hline
    0.5 & 190.841 & 1 & 90.314 \\
    \hline
    1 & 145.102 & 2 & 61.261\\
    \hline
    \end{tabular}
    
    \label{tab3}
\end{table}

\section{Computation of conductivity}
\noindent Now we will compute the conductivity in four dimension as a function of frequency, that is optical conductivity. In this section we work with $\psi_{-}=\langle \mathcal{O}_{\Delta_-}\rangle$ and $\psi_{+}=0$. We apply a sinusoidal electromagnetic perturbation in the bulk of frequency $\omega$ in $x$-direction. By the gauge/gravity duality, the fluctuation in the Maxwell field in the bulk gives rise to the conductivity.

\noindent Making the ansatz $A_{\mu}=(0,0,\varphi(r,t),0)$ with $\varphi(r,t)=A(r)e^{-i\omega t}$ leads to the following equation of motion for $A(r)$
\begin{eqnarray}
A^{\prime\prime}(r)+\frac{f^{\prime}(r)}{f(r)}A^{\prime}(r)+\left[\frac{\omega^2}{f^2(r)}-\frac{2\psi^2(r)}{f(r)} \right ]A(r)=0.
\label{50}
\end{eqnarray}
\noindent For conductivity calculation, we use the metric $f(z)\approx\frac{r_{+}^2}{z^2}g(z)$.

\noindent We now move in tortoise coordinate which is defined by

\begin{eqnarray}
r_{\star}&=&\int \frac{dr}{f(r)}=-\frac{1}{r_{+}}\int \frac{dz}{(1-z^3)\left[1+m_{g}^2\left \{ \frac{\bar{c_1}}{2} \frac{z(1-z^2)}{1-z^3}+\bar{c_2}\frac{z^2(1-z)}{1-z^3}\right \} \right ]}\nonumber\\
&\approx & -\frac{1}{r_{+}}\int \frac{dz}{1-z^3}\left [ 1-\frac{m_{g}^2\bar{c_1}}{2}\frac{z(1-z^2)}{1-z^3}-m_{g}^2\bar{c_2}\frac{z^2(1-z)}{1-z^3} \right ]\nonumber\\
&=&\frac{1}{6r_{+}}\Bigg [ \left \{ 1-\frac{m_{g}^2}{3}(\bar{c_{1}}+\bar{c_2}) \right \}ln\frac{(1-z)^3}{1-z^3}-2\sqrt{3}\left ( 1-\frac{m_{g}^2}{3}\bar{c_2} \right )tan^{-1} \frac{\sqrt{3}z}{2+z} \nonumber\\
&-& m_{g}^2\frac{z(1-z)}{1-z^3}\left \{ \bar{c_1}+2\bar{c_2}(1+z) \right \} \Bigg ].
\label{51}
\end{eqnarray}

\noindent   The integration constant has been so chosen such that $r_{\star}(z)=0$ at $z=0$. Eq.(\ref{50}) in tortoise coordinate reads

\begin{equation}
\frac{d^2A}{dr_{\star}^2}+\left [ \omega^2-V\right ]A=0~,~~~V=2\psi^2f.
\label{52}
\end{equation}

\noindent We shall replace the potential $V$ with its average $\left \langle \ V \right \rangle$ in a self-consistent manner \cite{siop}. With this approximation, The solution of eq.(\ref{52}) reads

\begin{equation}
A(z)\sim e^{-i\sqrt{\omega^2-\left \langle V \right \rangle}~r_{\star}(z)}.
\label{53}
\end{equation}

\noindent To obtain conductivity, we now expand the gauge field about $z=0$

\begin{equation}
A(z)=A(0)+ z A^{\prime}(0)+\mathcal{O}(z^2).
\label{54}
\end{equation}

\noindent Now in general $A_{x}$ can be written as

\begin{equation}
A_{x}=A_{x}^{(0)}+\frac{A_{x}^{(1)}}{r_{+}}z+\mathcal{O}(z^2).
\label{55}
\end{equation}

\noindent Now from the definition of conductivity and gauge/gravity correspondence, we have

\begin{equation}
\sigma(\omega)=\frac{\left \langle J_{x} \right \rangle}{E_{x}}=\frac{A_{x}^{(1)}}{-\dot{A_{x}}^{(0)}}=-\frac{r_{+}A^{\prime}(z=0)e^{-i \omega t}}{\frac{d}{dt}\left\{A(z=0)e^{-i \omega t}\right\}}=-\frac{i}{\omega}\frac{r_{+}A^{\prime}(z=0)}{A(z=0)}.
\label{56}
\end{equation}

\noindent By putting the approximate form of $A(z)$ in eq.(\ref{56}), we deduce the expression of conductivity

\begin{equation}
\sigma(\omega)=\sqrt{1-\frac{\left \langle V \right \rangle}{\omega^2}}~.
\label{57}
\end{equation}

\noindent We will now calculate the average value of the potential. This reads 

\begin{equation}
\left \langle V \right \rangle=\frac{\int_{-\infty}^{0}dr_{\star}V A^{2}(r_{\star})}{\int_{-\infty}^{0}dr_{\star} A^{2}(r_{\star})}.
\label{58}
\end{equation}

\noindent From eq.(\ref{52}) and $\psi(z)=\frac{\left \langle \mathcal{O}_{\Delta_-}\right \rangle}{\sqrt{2}r_{+}^{\Delta_-}}z^{\Delta_-}F(z)$, the potential reads 

\begin{equation}
V(z)=\frac{{\left \langle \mathcal{O}_{\Delta_-} \right \rangle}^2}{r_{+}^{2(\Delta_- -1)}}z^{2(\Delta_- -1)}F^2(z)\left [ g_{0}(z)+g_{1}(z) \right ].
\label{59}
\end{equation}

\noindent When $r_{+}\rightarrow 0$, the main contribution to the integrals in eq.(\ref{58}) is from the vicinity of the boundary where $r_{\star}\approx-\frac{z}{r_{+}}$. This is the interesting fact that  the leading order contribution in the nature of $r_{\star}$ is independent of mass of graviton or coupling parameters. We deduce the leading contribution to be 

\begin{equation}
\left \langle V \right \rangle \approx \frac{{\left \langle \mathcal{O}_{\Delta_-}\right \rangle }^2}{r_{+}^{2(\Delta_- -1)}}\left \{ \frac{\int_{0}^{\infty}dz z^{2(\Delta_- -1)}e^{2i\sqrt{\omega^2-\left \langle V \right \rangle}\frac{z}{r_{+}}}}{\int_{0}^{\infty}dz e^{2i\sqrt{\omega^2-\left \langle V \right \rangle}\frac{z}{r_{+}}}} \right \}={\left \langle \mathcal{O}_{\Delta_-} \right \rangle }^2\left [ \frac{\Gamma (2\Delta_- -1)}{\left \{ -2i\sqrt{\omega^2-\left \langle V \right \rangle } \right \}^{2(\Delta_- -1)}} \right ].
\label{60}
\end{equation}
\noindent which determines $\left \langle V \right \rangle$ implicitly as a function of $\omega$. The low temperature high frequency conductivity therefore reads

\begin{equation}
\sigma(\omega)=\sqrt{1-\frac{\Gamma(2\Delta_--1)(-2i)^{2(1-\Delta_-)}{\left \langle \mathcal{O}_{\Delta_-} \right \rangle }^2}{\omega^{2\Delta_-}}}
\label{61}
\end{equation}
 whereas for the low frequency, we have
 \begin{equation}
 \sigma(\omega)=\sqrt{1-\frac{\left[2^{2(1-\Delta_-)}\Gamma(2\Delta_--1){\left \langle \mathcal{O}_{\Delta_-} \right \rangle }^2\right]^{\frac{1}{\Delta_-}}}{\omega^2}}.
 \label{62}
 \end{equation}
 
 \noindent In particular for $\Delta_-=1$, eq.(\ref{60}) can be solved for all frequencies and the expression (\ref{61}) for the conductivity, which coincides with eq.(\ref{62}), is valid in the entire spectrum
 \begin{equation}
 \sigma(\omega)=\sqrt{1-\frac{{\left \langle \mathcal{O}_{1} \right \rangle }^2}{\omega^2}}=\frac{i\left \langle \mathcal{O}_{1} \right \rangle }{\omega}\sqrt{1-\frac{\omega^2}{{\left \langle \mathcal{O}_{1} \right \rangle}^2 }}~.
 \label{63}
 \end{equation}
 
 \noindent But due to the approximation of replacing potential by its self-consistent average, we can not see the effect of mass of graviton and coupling parameters in the expression of conductivity. We just get the leading order contribution term of conductivity by this method. 
 
\noindent We now first solve eq.(\ref{52}) for near horizon region that is $V\approx0$. The solution reads

\begin{equation}
A \sim e^{-i\omega r_{\star} }\sim (1-z)^{-\frac{i\omega}{3 r_{+}}\left \{ \frac{1}{1+\frac{1}{3}m_{g}^2(\bar{c_{1}}+\bar{c_{2}}) }\right \}}
\label{64}
\end{equation}
 where we consider only leading order term in $r_{\star}$, that is $r_{\star}=\frac{1}{3 r_{+}}\left \{ 1-\frac{1}{3}m_{g}^2(\bar{c_{1}}+\bar{c_{2}}) \right \}ln(1-z)$ in the obtaining above expression. Now we write eq.(\ref{50}) in $z$-coordinate. This reads
 
 \begin{equation}
 g(z)\frac{d^2A(z)}{dz^2}+g^{\prime}(z)\frac{dA(z)}{dz}+\left [ \frac{\omega^2}{r_{+}^2g(z)}-\frac{2\psi^2(z)}{z^2} \right ]A(z)=0~.
\label{65}
 \end{equation}
 
 \noindent We now write $A(z)$ as a product of $\omega$-dependent part and a function of $z$ which we need to determine. Hence the gauge field reads
 \begin{equation}
 A(z)=(1-z)^{-\frac{i \omega}{3 r_{+}}C_{1}}G(z),~~~C_{1}=\frac{1}{ 1+\frac{1}{3}m_{g}^2(\bar{c_{1}}+\bar{c_{2}})}
 \label{66}
 \end{equation}
 where $G(z)$ is regular at the horizon of black hole.
 
 \noindent Substituting the above expression for $A(z)$ in eq.(\ref{65}), we obtain
 
 \begin{eqnarray}
 g(z)\frac{d^2G(z)}{dz^2}+\left [\frac{2i\omega C_{1}}{3r_{+}}\frac{g(z)}{1-z}+g^{\prime}(z)\right ]\frac{dG(z)}{dz}+\Bigg [\frac{i\omega C_{1}}{3r_{+}}\left (  \frac{i\omega C_{1}}{3r_{+}}+1\right )\frac{G(z)}{(1-z)^2}\nonumber\\
 +\frac{i\omega C_{1}}{3r_{+}}\frac{G^{\prime}(z)}{1-z}+\frac{\omega^2}{r_{+}^2g(z)}-\frac{2\psi^2(z)}{z^2} \Bigg ]G(z)=0.
 \label{67}
 \end{eqnarray}
 
 \noindent For $\Delta_-=1$ that is $m^2=-2$, we know that $\psi(z)\approx\frac{\left \langle \mathcal{O}_{1} \right \rangle }{\sqrt{2}r_{+}}z$ where. Substituting this in eq.(\ref{67}), we get 
 \begin{eqnarray}
 3g_{0}(z)\frac{d^2G(z)}{dz^2}&+&\left [\frac{2i\omega C_{1}}{r_{+}}(1+z+z^2)-9z^2C_{2}(z)\right ]\frac{dG(z)}{dz}+\Bigg [\frac{i\omega C_{1}}{r_{+}}\frac{\left \{  1+z+z^2-3z^2C_{2}(z) \right \}}{1-z}\nonumber\\
 &+& \frac{\omega^2}{3r_{+}^2}\left \{ \frac{9C_{3}(z)-C_{1}^2(1+z+z^2)^2}{1-z^3}\right \} -\frac{3\left \langle \mathcal{O}_{1}\right \rangle^2}{r_{+}^2}C_{4}(z)\Bigg ]G(z)=0
 \label{68}
 \end{eqnarray}
 where
 
 \begin{equation}
 C_{2}(z)=\frac{1+\frac{g_{1}^{\prime}(z)}{g_{0}^{\prime}(z)}}{1+\frac{g_{1}(z)}{g_{0}(z)}};~~~C_{3}(z)=\frac{1}{\left[1+\frac{g_{1}(z)}{g_{0}(z)}\right]^2};~~~~C_{4}(z)=\frac{1}{1+\frac{g_{1}(z)}{g_{0}(z)}}.
 \label{69}
 \end{equation}
 \noindent After some long algebra, we obtain
 \begin{eqnarray}
 3g_{0}(z)\frac{d^2G(z)}{dz^2}+\left [\frac{2i\omega C_{1}}{r_{+}}(1+z+z^2)-9z^2C_{2}(z)\right ]\frac{dG(z)}{dz}+\Bigg [\frac{i\omega C_{1}}{r_{+}}\left \{ 1+2z+\frac{3z^2[1-C_{2}(z)]}{1-z} \right \}\nonumber\\
 +\frac{\omega^2}{3r_{+}^2}\left \{ \frac{[9C_{3}(z)-C_{1}^2](1+z+z^2+z^3)-C_{1}^2z(2+5z+7z^2)+\frac{9z^4\left [ C_{3}(z)-C_{1}^2 \right ]}{1-z}}{1+z+z^2}\right \}\nonumber\\
  -\frac{3\left \langle \mathcal{O}_{1} \right \rangle^2}{r_{+}^2}C_{4}(z)\Bigg ]G(z)=0.
  \label{70}
 \end{eqnarray}
 \noindent At the horizon we now expand $G(z)$ in a Taylor series. We deduce the boundary condition
 
 \begin{equation}
 \left(3-\frac{2i\omega C_{1}}{r_{+}}\right)G^{\prime}(1)+\left[\frac{\left \langle \mathcal{O}_{1}\right \rangle^2 C_{1}}{r_{+}^2}-\frac{i\omega}{3 r_{+}}\left\{3C_{1}(1+t)-\frac{i\omega}{r_{+}}(2C_{1}^2+s)\right\}\right]G(1)=0
 \label{71}
 \end{equation}
 where
 \begin{equation}
 t=\frac{m_{g}^2(\bar{c_{1}}+2\bar{c_{2}})}{6\left\{1+\frac{m_{g}^2}{3}(\bar{c_{1}}+\bar{c_{2}})\right\}};~~~s=\frac{m_{g}^2(\bar{c_{1}}+2\bar{c_{2}})}{3\left\{1+\frac{m_{g}^2}{3}(\bar{c_{1}}+\bar{c_{2}})\right\}^3}
 \label{72}
 \end{equation}
 and we use the fact that
 \begin{equation}
 C_{2}(1)=1;~~~ C_{1}=\sqrt{C_{3}(1)}=C_{4}(1).
 \label{73}
 \end{equation}

 \noindent To solve eq.(\ref{70}) for low frequency region ($\omega<<\left \langle \mathcal{O}_1 \right \rangle$), we rescale it by letting $z=\frac{z^{\prime}}{a}$, where $a=\frac{\left \langle \mathcal{O}_{1} \right \rangle}{r_{+}}$ and then take the $a\rightarrow \infty$  which correspond to the low temperature regime. This leads to
 \begin{equation}
 \frac{d^2G(z^{\prime})}{dz^{\prime^2}}-G(z^{\prime})=0.
 \label{74}
 \end{equation}
 
 \noindent The solution of this equation reads
 
 \begin{eqnarray}
 G(z^{\prime})&=&C_{+}e^{z^{\prime}}+C_{-}e^{-z^{\prime}}\nonumber\\
\implies G(z)&=&C_{+}e^{az}+C_{-}e^{-az}\nonumber\\
&=&C_{+}e^{\frac{\left \langle \mathcal{O}_{1} \right \rangle}{r_{+}}z}+C_{-}e^{-\frac{\left \langle \mathcal{O}_{1} \right \rangle}{r_{+}}z}.
\label{75}
 \end{eqnarray}
 
 \noindent The approximate solution near the boundary is found to be
 \begin{equation}
 A(z)=e^{\frac{i\omega z}{3 r_{+}}\left \{ \frac{1}{1+\frac{1}{3}m_{g}^2(\bar{c_{1}}+\bar{c_{2}}) }\right \}}\left[C_{+}e^{\frac{\left \langle \mathcal{O}_{1} \right \rangle}{r_{+}}z}+C_{-}e^{-\frac{\left \langle \mathcal{O}_{1} \right \rangle}{r_{+}}z}\right].
\label{76}
\end{equation}  

\noindent By putting the upper form of $A(z)$ in the eq.(\ref{56}), we obtain

\begin{equation}
\sigma(\omega)\approx i\frac{1-\frac{C_{+}}{C_{-}}}{1+\frac{C_{+}}{c_{-}}}\frac{\left \langle \mathcal{O}_{1} \right \rangle}{\omega}.
\label{77}
\end{equation} 

\noindent The ratio $C_{+}/C_{-}$ is found by substituting $G(1)$ and $G^{\prime}(1)$ from eq.(\ref{76}) in the boundary condition (\ref{71}). This yields up to first order in $\omega$ 

\begin{equation}
\frac{C_{+}}{C_{-}}=-e^{-2a}\left[\frac{aC_{1}-3}{aC_{1}+3}+\frac{2\left\{2a^2C_{1}-3(1+t)\right\}}{a(aC_{1}+3)^2}\frac{iC_{1}\omega}{r_{+}}+\mathcal{O}(\omega^2)\right],~~~a=\frac{\left \langle \mathcal{O}_{1} \right \rangle}{r_{+}}.
\label{78}
\end{equation}

\noindent We deduce the expression of low frequency ($\omega<<\left\langle \mathcal{O}_{1}\right \rangle$) conductivity at low temperature as

\begin{equation}
\sigma(\omega)=\frac{i\left \langle \mathcal{O}_{1}\right \rangle}{\omega}\left[1+2\frac{aC_{1}-3}{aC_{1}+3}e^{-2a}+4\frac{\left\{2a^2C_{1}-3(1+t)\right\}}{a(aC_{1}+3)^2}e^{-2a}\frac{iC_{1}\omega}{r_{+}}+\mathcal{O}(\omega^2)\right]~.
\label{79}
\end{equation}
This is in agreement with the leading order result (\ref{63}). The above result is useful in estimating the band gap energy of the holographic superconductor. The DC conductivity is 
 
 \begin{eqnarray}
 Re~ \sigma(\omega=0)=-8\left\{1-\frac{3(1+t)}{2a^2C_{1}}\right\}\left\{1+\frac{3}{aC_{1}}\right\}^{-2}e^{-2a}\nonumber\\
 \sim \left [ 1+\mathcal{O}\left ( \frac{1}{a} \right ) \right ]e^{-2a}\approx e^{-2\frac{\left \langle \mathcal{O}_{1} \right \rangle}{r_{+}}}=e^{-\frac{E_g}{T}}
 \label{80}
 \end{eqnarray}
where 

\begin{eqnarray}
E_{g}=\frac{3}{2\pi}\left\{1+\frac{m_{g}^2}{3}\left(\bar{c_1}+\bar{c_2}\right)\right\}\left \langle \mathcal{O}_{1}\right \rangle.
\label{81}
\end{eqnarray} 

\noindent $E_g$ is identified to be the band gap energy. We recover the band gap energy $E_g=\frac{3\left\langle \mathcal{O}_{1}\right \rangle}{2\pi}\approx 0.48\left\langle \mathcal{O}_{1}\right \rangle$ \cite{siop} when mass of graviton is zero. The result indicates that $E_g$ increases with increase in the mass of the graviton and coupling parameters.

\section{Conclusions}
In this paper, based on Sturm-Liouville eigenvalue problem, we perform analytic computation of holographic $s$-wave superconductor in massive gravity background. We found that the massive gravity parameters affect significantly on critical temperature and condensate value. When increasing the massive gravity couplings, the critical temperature increases, whereas the condensate value gets lower. It is further observed that above a critical value of each massive gravity coupling, the critical temperature in massive gravity is higher than that in Einstein gravity, whereas the condensate value is smaller compared to the case of massless graviton. Then we have analytically computed the conductivity of holographic superconductor in the massive gravity backgound in four dimensions. First we perform the computation of conductivity by following a self consistant approach. Then we obtain the expression of conductivity that contains effect of coupling parameters and mass of graviton and finally we found that leading order results obtained from these two approaches are consistent. From the real part of the DC conductivity, the band gap energy is obtained. We found that band gap energy increases with the increasing values of mass of graviton and coupling parameters.
These facts suggest that the high temperature superconductors can be achieved in the presence of graviton mass.
\section*{Acknowledgments}
DP and SP would like to thank CSIR for financial support.


\begin{thebibliography}{99}
\baselineskip=0.6 cm
\bibitem{adscft1} J. M. Maldacena, Adv. Theor. Math. Phys. 2 (1998) 231.
\bibitem{adscft2} E. Witten, Adv. Theor. Math. Phys. 2  (1998) 253.
\bibitem{adscft3} S.S. Gubser, I.R. Klebanov, A.M. Polyakov, Phys. Lett. B 428 (1998) 105.
\bibitem{adscft4} O.Aharony, S.S. Gubser, J.M. Maldacena, H.Ooguri, Y.Oz.,Phys. Rep.,323(2000)183.
\bibitem{adscft5} S.A. Hartnoll, C.P. Herzog, G.T. Horowitz, Phys. Rev. Lett. 101 (2008) 031601.
\bibitem{ssg} S.S. Gubser, Phys. Rev. D 78 (2008) 065034.

\bibitem{hs9} S. A. Hartnoll, C. P. Herzog, G. T. Horowitz, JHEP 12 (2008) 015.


\bibitem{hs8} R. Gregory, S. Kanno, J. Soda, JHEP 0910 (2009) 010.
\bibitem{siop} G. Siopsis, J. Therrien, JHEP 05 (2010) 013.
\bibitem{hs17} G. T. Horowitz, M. M. Roberts, JHEP  0911 (2009) 015.
\bibitem{gm} G. T. Horowitz, M. M. Roberts, Phys. Rev. D 78, 126008 (2008).
\bibitem{hs9a} H.B. Zeng, X. Gao, Y. Jiang, H.-S. Zong, JHEP 05 (2011) 002.
\bibitem{hs9b} H.F. Li, R.-G. Cai, H.-Q. Zhang, JHEP 04 (2011) 028.
\bibitem{hs14}  Q. Y. Pan, B. Wang, E. Papantonopoulos, J. Oliveira, A. Pavan, Phys. Rev. D 81  (2010) 106007.
\bibitem{hs15}  R.G. Cai, H. Zhang, Phys. Rev. D 81 (2010) 066003.
\bibitem{horowitz} G. T. Horowitz and B. Way, J. High Energy Phys. 11
(2010) 011.
\bibitem{hs19} J. Jing, S. Chen, Phys. Lett. B 686 (2010) 68.
\bibitem{hs20} J. Jing, Q Pan, S. Chen, JHEP 1111 (2011) 045. 
\bibitem{sgdc1} S.~Gangopadhyay, D.~Roychowdhury, JHEP 05 (2012) 002.
\bibitem{hs24} S.~Gangopadhyay, D.~Roychowdhury, JHEP 05 (2012) 156.
\bibitem{sg-single}S.~Gangopadhyay, Phys. Lett. B 724 (2013) 176.
\bibitem{rb} R. Banerjee, S.~Gangopadhyay, D.~Roychowdhury, A. Lala, Phys. Rev. D 87 (2013) 104001.
\bibitem{sgm} S. Gangopadhyay, Mod. Phys. Lett. A 29 (2014) 1450088.
\bibitem{dg3} D.~Ghorai, S.~Gangopadhyay, Eur. Phys. J. C 76 (2016) 702.
\bibitem{dg1} D. Ghorai, S. Gangopadhyay, Eur. Phys. J. C 76 (2016) 146. 
\bibitem{dg4} D.~Ghorai, S.~Gangopadhyay,  Euro. Phys. Lett. 118 (2017) 31001.
\bibitem{dg6} D.~Ghorai, S.~Gangopadhyay, Nucl. Phys. B933 (2018) 1-13.
\bibitem{dpdgsg} D.~Prai, D.~Ghorai, S.~Gangopadhyay, General Relativity and Gravitation, 50(11), 1-17.
\bibitem{su} S.Pal, S. Gangopadhyay, Ann. Phys., 388 (2018), p. 472.
\bibitem{sup} S. Pal, S. Ghosh,  S. Gangopadhyay, Ann. Phys. 414 (2020), 168078.
\bibitem{ss} S.S. Gubser, S.S. Pufu, J. High Energy Phys. 11 (2008) 033.
\bibitem{s} S.S. Gubser. Phys. Rev. Lett. 101 (2008) 191601.
\bibitem{sd} S. Gangopadhyay, D. Roychowdhury, J.High Energy Phys.08 (2012) 104.
\bibitem{spdg} S. Pal, D. Parai, S. Gangopadhyay, Phys. Rev. D, 105(6), 065015. 

\bibitem{dg1} D. Parai, D.Ghorai,  S. Gangopadhyay, Int. J. Mod Phys. A 37, no. 1 (2022): 2150249.
\bibitem{dg2} D. Parai, D.Ghorai, S. Gangopadhyay, Ann. Phys, 403, 59-67.
\bibitem{dg3} D. Parai, D.Ghorai, S. Gangopadhyay, The Eur. Phys. J. C, 80(3), 1-9.

\bibitem{mw} M. Fierz, W. Pauli, Proc. R. Soc. A 173 (1939) 211.
\bibitem{hm} H. Van Dam, M.J.G. Veltman, Nucl. Phys. B 22 (1970) 397.
\bibitem{vp} V.I. Zakharov, Pis'ma Zh. Eksp. Teor. Fiz. 12 (1970) 447, JETP Lett. 12 (1970) 312.
\bibitem{a} A.I. Vainshtein, Phys. Lett. B 39 (1972) 393.
\bibitem{ds} D.G. Boulware, S. Deser, Phys. Rev. D 6 (1972) 3368.
\bibitem{cg} C de Rham, G. Gabadadze, Phys. Rev. D 82 (2010) 044020.
\bibitem{cga} C de Rham, G. Gabadadze, A.J. Tolley, Phys. Rev. Lett. 106 (2011) 231101.
\bibitem{hj} H.B. Zeng, J.P.Wu, Phys. Rev. D 90 (2014) 046001.
\bibitem{ry} R. Li, Y. Zhao, Phys. Rev. D 100 (2019) 046018.







\bibitem{v} D. Vegh, arxiv:1301.0537
\bibitem{CHNam} C.H.Nam, Phys. Lett. B 807 (2020) 135547

\end{thebibliography}
\end{document}